\pdfoutput=1
\documentclass[10.5 pt]{article}
\usepackage{graphicx}
\usepackage{amssymb}
\usepackage{amsmath}
\usepackage{amsfonts}
\usepackage{amsthm}
\usepackage[english]{babel}
\usepackage[latin1,ansinew]{inputenc}
\usepackage{a4wide}
\usepackage{color}
\newcommand{\be}{\begin{eqnarray}}
\newcommand{\ee}{\end{eqnarray}}

\begin{document}

\title{\textbf{On the Convergence of Strong Cylindrical and\\ Spherical Shock Waves in Solid Materials}}
\author{\textbf{R. K. Anand} 
\footnotemark[1]}


\maketitle

\footnotetext[1]{Department of Physics, UGC Centre of Advanced Studies, University of Allahabad, Prayagraj 211002, India \\ email: rkanand@allduniv.ac.in; rajkumaranand@rediffmail.com\\ Accepted in: Proc. Natl. Acad. Sci., India, Sect. A Phys. Sci. 2025 DOI : 10.1007/s40010-025-00909-y}
%

\begin{abstract}
In this article, we present a description of the behaviour of shock-compressed solid materials following Geometrical Shock Dynamics (GSD) theory. GSD has been successfully applied to various gas dynamics problems, and here we have employed it to investigate the propagation of cylindrically and spherically symmetric converging shock waves in solid materials. The analytical solution of shock dynamics equations has been obtained in strong-shock limit, assuming the solid material to be homogeneous and isotropic and obeying the Mie-Gr\"uneisen equation of state. The non-dimensional expressions are obtained for the velocity of shock, the pressure, the mass density, the particle velocity, the temperature, the speed of sound, the adiabatic bulk modulus, and the change-in-entropy behind the strong converging shock front. The influences as a result of changes in (i) the propagation distance $r$ from the axis or centre $(r=0)$ of convergence, (ii)  the Gr\"uneisen parameter, and (iii)  the material parameter are explored on the shock velocity and the domain behind the converging shock front. The results show that as the shock focuses at the axis or origin, the shock velocity, the pressure, the temperature, and the change-in-entropy increase in the shock-compressed titanium Ti6Al4V, stainless steel 304, aluminum 6061-T6, etc.
\end{abstract}

\textbf{Relevance of research} 
{The study of converging shock in solids is relevant to the production of very high pressure and temperature in condensed materials. Such studies yield information on the equation of state of solids subjected to high pressures,  which is very important for solving a large number of problems in applied physics, engineering, astrophysics, geophysics, material science, and other branches of science. The present article discloses the behaviour of shock-compressed titanium Ti6Al4V, brass (66\% copper and 34\% zinc), tantalum, iron, stainless steel 304, aluminum 6061-T6, and OFHC copper.}

\section{Introduction}
\label{intro}

The study of the behaviour of shock-compressed materials plays an important role in the aerospace, ballistic, and industrial applications of solids. As the converging shock approaches the axis or centre, it strengthens, and the shock speed, pressure, and temperature increase rapidly. In 1942, Guderley \cite{Guderley1942} first studied converging shock waves in an inviscid perfect gas. He predicted that the shock strength varies inversely with a power of propagation distance, which means that the pressures and temperatures will be extreme at the centre of convergence. However, the first experimental study on converging shock waves was done in 1951 by Perry and Kantrowitz \cite{Perry1951} at 1.4 and 1.8 shock Mach numbers. The Geometrical Shock Dynamics (GSD) theory developed by Chester \cite{Chester1954}, Chisnell \cite{Chisnell1957}, and Whitham \cite{Whitham1958} is a simple and useful theoretical tool to analyse the process of propagation of shock waves in uniform and non-uniform media. For a detailed explanation of the use of GSD theory, the reader is referred to Refs. \cite{Apazidis2019,Han1992,Henshaw1986}. Recently, Ndebele and Skews \cite{Ndebele2019} have revisited Guderley's problem using GSD theory and shown good approximations to the values of the Guderley exponent. In general, GSD theory is being applied to a wide variety of engineering and scientific applications \cite{Peton2020,Peace2018,Ramsey2012} as it provides extremely precise results, particularly for spherically symmetric shocks \cite{Ridoux2015,Schwendeman1993,Best1991}. Using GSD theory, shock waves in water and in gases with complex geometries were also investigated by Cates and Sturtevant \cite{Cates1997} and Aslam and Stewart \cite{Aslam1999}, respectively. Madhumita and Sharma \cite{Madhumita2003} extended the problem of strong converging cylindrical and spherical shock waves in an inhomogeneous gaseous medium using a perturbative approach. Anand \cite{Anand2013a,Anand2013b} has also used GSD theory for analyzing imploding shock waves in non-ideal gases. Ridoux et al. \cite{Ridoux2020} investigated the propagation of blast waves in the presence of obstacles employing GSD theory and proposed a fast-running method. Singh and Arora studied the cylindrical shock waves in a non-ideal gas \cite{Singh2020a,Singh2020b} and further extended the same work for converging cylindrical shock \cite{Singh2020b}, taking into account the presence of a magnetic field \cite{Singh2020c} and dust particles \cite{Singh2022}. Recently, Anand and Singh studied the shock waves in van der Waals gases and investigated the effects of gravitational and magnetic fields on the propagation of cylindrical strong \cite{Anand2024a} and the viscosity on the structure of shock waves \cite{Anand2024b}. 

Ramsey et al. \cite{Ramsey2018}, Boyd et al. \cite{Boyd2017}, Kanel \cite{Kanel2004} and many others have shown that the high pressures and temperatures inside the shock-compressed solids may change their crystal structure, melting, vaporisation, and other material properties. The  Mie-Gr\"uneisen equation of state \cite{Mie1903,Gruneisen1912} is frequently used to investigate the behaviour of shock-compressed solids materials. The study of Yadav and Singh \cite{Yadav1982} on the converging shock waves in metals showed that in strong-shock limit the light metals e.g. aluminum behave similar to gases  while heavier metals e.g. copper show a little variation of this behaviour. Using Mie-Gr\"uneisen equation of state Lieberthal et al. \cite{Lieberthal2017}, L\'{}opez et al. \cite{Lopez2011}, Arienti et al. \cite{Arienti2004} and many others have employed different methods and techniques to study the motion of shock waves in solid materials. Anand \cite{Anand2022} has applied GSD theory to present a model for weak converging shock waves in solid materials. Recently, Anand and Singh \cite{Anand2023} investigated shock waves in tungsten and titanium metals by using the linear theory of the viscous stress tensor.

The study of converging shock in solids is relevant to the production of very high pressure and temperature in condensed materials. Such studies yield information on the equation of state of solids subjected to high pressures, which is very important for solving a large number of problems in applied physics, engineering, astrophysics, geophysics, material science, and other branches of science. The present article discloses the behaviour of shock-compressed titanium Ti6Al4V, brass (66\% copper and 34\% zinc), tantalum, iron, stainless steel 304, aluminum 6061-T6, and OFHC copper.

The purpose of present study is to disclose the behaviour of shocked materials like  aluminum, brass (66\% copper and 34\% zinc), copper and stainless steel. The GSD theory works well with the converging shock waves, and the results obtained are in good approximations with experiments \cite{Hornung2008,Schwendeman1987}. Therefore, GSD theory is applied for investigating the propagation of cylindrically and spherically symmetric converging shock waves in solid materials. 
The analytical solution of shock dynamics equations has been obtained in strong-shock limit, assuming that
(i) the solid material behaves like a fluid at the instant when shock passes through it and obeys Mie-Gr\"uneisen equation of state, (ii) the shocked material is homogeneous, isotropic, and chemically nonreactive, (iii) the interactions among constituents are negligible, and (iv) the shock wave propagates steadily with a step wave profile, i.e., with zero-rise time. These solutions are generally applicable to Mach numbers greater than 2.
The non-dimensional expressions behind the strong converging shock front are obtained for the velocity of shock, the pressure, the mass density, the particle velocity, the temperature, the speed of sound, the adiabatic bulk modulus, and the change-in-entropy in shocked material under the equilibrium condition. The influences as a result of changes in (i) the propagation distance $r$ from the axis or centre $(r=0)$ of convergence, (ii) the Gr\"uneisen parameter, and (iii) the material parameter are explored on the shock velocity and the domain behind the converging shock-front. The results show that as the shock approaches the axis or origin, the shock velocity, the pressure, the temperature and, the change-in-entropy increase in the shock-compressed titanium Ti6Al4V, OFHC copper, stainless steel 304, aluminum 6061-T6, etc.

The remainder of the paper is coordinated as follows: In  Sect. \ref{form}, the formulation of the problem is presented.
In Sect. \ref{gsd}, we describe how GSD theory is used for solid materials. The effects on the thermodynamic properties of shocked materials such as pressure, temperature, mass density, and entropy are discussed in detail and compared with the experimental observations in Sect. \ref{res}. 

\section{Formulation of problem}
\label{form}
Let us consider a 1-D shock wave propagating in a uniform and isotropic solid material in an equilibrium state. The conservation equations governing the symmetric motion of the shock wave can be expressed conveniently in Eulerian coordinates as:

\begin{equation}\label{eq-1}
\frac{\partial u}{\partial t} + u\frac{\partial u}{\partial r}+\frac{1}{\rho}\frac{\partial p}{\partial r}=0,
\end{equation}

\begin{equation} \label{eq-2}
\frac{\partial \rho}{\partial t} + u\frac{\partial \rho}{\partial r}+\rho\left(\frac{\partial u}{\partial r}+j\frac{u}{r}\right)=0,
\end{equation}

\begin{equation} \label{eq-3} 
\frac{\partial p}{\partial t} + u\frac{\partial p}{\partial r}+K_{s}\left(\frac{\partial u}{\partial r}+\frac{ju}{r}\right) =0,
\end{equation}

where $r$ is the position of the shock front from the axis or centre $(r=0)$ at time $t$. $u(r,t)$, $p(r,t)$ and $\rho(r,t)$ are the particle (bulk flow) velocity, the pressure (scalar), and the mass density, respectively. The geometrical factor $j$ is $0,1,$ and $2$, respectively, for the planar, cylindrical and spherical symmetry. $K_{s}(\rho,p)$, the adiabatic bulk modulus or inverse compressibility,  is defined by $K_{s}=\rho\left(\frac{\partial p}{\partial\rho}\right)_{s}$, where $s$ is the entropy of material.

Ramsey et al. \cite{Ramsey2018}, Boyd et al. \cite{Boyd2017}, Lieberthal and Stewart \cite{Lieberthal2017}, L\'{}opez et al. \cite{Lopez2011}, Arienti et al. \cite{Arienti2004} and many others have considered the Mie-Gr\"{u}neisen equation of state as: $p=e\rho\Gamma$, where $e$ is the internal energy per unit mass or specific internal energy and $\Gamma$ is the Gr\"{u}neisen coefficient, for investigating the motion of shock waves in the solids with sufficient accuracy. Bushman and Fortov \cite{Bushman1983} and Anisimov and Kravchenko \cite{Anisimov1985} have considered the Gr\"{u}neisen coefficient as: 

\begin{equation}\label{eq-4}
\Gamma\left(G\right)=\frac{3}{2}+\left(\Gamma_{o}-\frac{2}{3}\right)\frac{G_{m} ^{2}+1}{G_{m} ^{2}+G^{2}}G,
\end{equation}
where $\Gamma_{o}$, $G_{m}$, and $G$ are the Gr\"uneisen parameter, the material parameter, and the shock-compression ratio. $\Gamma_{o}$ and $G_{m}$ are determined, in general, by the experiments. The expression $K_{s}=\rho a^{2}$ connects $K_s$ to $a$. Here $a$, the local speed of sound, is equal to $\sqrt{(\Gamma+1)p/\rho}$.\\

\begin{table}
\caption{Computed values of the parameter $G$}
\label{tab:1}
\begin{tabular}
{p{1.0 cm} p{1.5 cm} p{1.5 cm} p{1.5 cm} p{1.5 cm} p{1.5 cm} p{1.5 cm} p{1.5 cm} p{1.5 cm}}
\hline\noalign{\smallskip}
$G_{m}$ & Ti & Brass & Ta  &  Fe & Steel & Al  & Cu \\
 \noalign{\smallskip}\hline\noalign{\smallskip}
0.51 & 3.27758 & 3.05567 & 2.81588 & 2.71579 & 2.58913 & 2.55723 & 2.51846 \\
0.53 & 3.26808 & 3.04413 & 2.80284 & 2.70237 & 2.57549 & 2.54358 & 2.50481 \\
0.55 & 3.25831 & 3.03230 & 2.78952 & 2.68870 & 2.56162 & 2.52971 & 2.49096 \\
0.57 & 3.24827 & 3.02018 & 2.77594 & 2.67479 & 2.54755 & 2.51565 & 2.47694 \\
0.59 & 3.23797 & 3.00779 & 2.76213 & 2.66068 & 2.53332 & 2.50144 & 2.46278 \\
0.61 & 3.22742 & 2.99516 & 2.74810 & 2.64638 & 2.51895 & 2.48710 & 2.44851 \\
0.63 & 3.21663 & 2.98228 & 2.73389 & 2.63192 & 2.50447 & 2.47266 & 2.43415 \\
0.65 & 3.20562 & 2.96919 & 2.71951 & 2.61733 & 2.48990 & 2.45815 & 2.41974 \\
0.67 & 3.19439 & 2.95590 & 2.70499 & 2.60264 & 2.47528 & 2.44360 & 2.40530 \\
0.69 & 3.18296 & 2.94242 & 2.69035 & 2.58786 & 2.46063 & 2.42903 & 2.39085 \\
0.71 & 3.17134 & 2.92878 & 2.67562 & 2.57303 & 2.44597 & 2.41446 & 2.37643 \\
0.73 & 3.15953 & 2.91499 & 2.66082 & 2.55816 & 2.43133 & 2.39993 & 2.36206 \\
0.75 & 3.14756 & 2.90107 & 2.64596 & 2.54329 & 2.41673 & 2.38546 & 2.34776 \\
0.77 & 3.13543 & 2.88704 & 2.63108 & 2.52843 & 2.40220 & 2.37106 & 2.33355 \\
0.79 & 3.12316 & 2.87292 & 2.61619 & 2.51360 & 2.38775 & 2.35676 & 2.31945 \\
 \noalign{\smallskip}\hline
\end{tabular}
\end{table}
At the shock front, $r=R(t)$, the boundary conditions for shock waves in solid materials are given by the Rankine-Hugoniot jump conditions \cite{Anand2022} as:
\begin{equation}\label{eq-5}
\frac{p}{p_{o}}=\frac{2\left(\Gamma+1\right)M^{2}}{\Gamma+2}-\frac{\Gamma}{\Gamma+2},
\end{equation}

\begin{equation}\label{eq-6}
\frac{\rho}{\rho_{o}}=\frac{\left(\Gamma+2\right)M^{2}}{\Gamma M^{2}+2},
\end{equation}

\begin{equation}\label{eq-7}
\frac{u}{a_{o}}=\frac{2}{\Gamma+2}\left(M-\frac{1}{M}\right),
\end{equation}

\begin{equation}\label{eq-8}
\frac{T}{T_{o}}=\frac{(2+\Gamma M^{2})\left[2\left(\Gamma+1\right)M^{2}-\Gamma\right]}{\Gamma\left(\Gamma +2\right)^{2} M^{2}},
\end{equation}  

\begin{equation}\label{eq-9}
\frac{a}{a_{o}}=\left[\frac{\left(\Gamma+1\right)\left(2+\Gamma M^{2}\right)\left[2M^{2}\left(\Gamma+1\right)-\Gamma\right]}{(\Gamma_{o}+1)(\Gamma+2)^{2}M^{2}}\right]^{1/2},
\end{equation} 
where $U=\frac{dR}{dt}$ is the shock velocity, $M = U/a_o$ is the shock Mach number, and $T$ is the absolute temperature of the shock-compressed material. The quantities just ahead of and behind the shock front are denoted by a subscript $o$ and without a subscript, respectively. The speed of sound $a_o$ in the unshocked material is equal to $\sqrt{(\Gamma_o+1)p_o/\rho_o}$. 
Using thermodynamic relations, the change-in-entropy $\Delta s/c_{v}$ across a shock-front of an arbitrary strength may be expressed as: $\Delta s/c_{v} = ln\left(p/p_o\right)-\left(\Gamma+1\right)ln\left(\rho / \rho_o\right)$. The jump relations across the strong shock waves ($p \gg p_o$) for the present problem can be written, using (\ref{eq-5})-(\ref{eq-9}), as:

\begin{equation}\label{eq-10}
p=\frac{2\rho_o(\Gamma+1)}{(\Gamma_o+1)(\Gamma+2)}U^2,
\end{equation}

\begin{equation}\label{eq-11}
\rho=\rho_o\frac{\Gamma+2}{\Gamma},
\end{equation}

\begin{equation}\label{eq-12}
u=\frac{2}{\Gamma+2}U,
\end{equation}

\begin{equation}\label{eq-13}
T=\frac{2\Gamma_o(\Gamma+1)T_o}{(\Gamma+2)^2}U^2,
\end{equation}

\begin{equation}\label{eq-14}
a=\frac{(\Gamma+1)\sqrt{2\Gamma(\Gamma_o+1)}}{(\Gamma_o+1)(\Gamma+2)}U.
\end{equation}

\section{Geometrical shock dynamics for solid materials}
\label{gsd}
According to GSD theory, the interactions of the converging shock front with the flow behind it are ignored as the shock is adjusting to changes in its geometry. The motion of the converging shock can be estimated by integrating the governing flow equations along the negative characteristic $dr/dt=u-a$ labeled by $C_-$. When the shock wave gets stronger, the slope of the $C_-$ curve family is close to the trajectory slope of a converging shock wave. To obtain the motion of a shock front, Whitham's characteristic rule evaluates characteristic $C_-$ using the flow states estimated from the Rankine-Hugoniot jump conditions. This step gives a relationship between the propagation distance $r$ and the shock velocity $U$, which is used to compute the motion of the shock front. 
The characteristic form of Eqs. (\ref{eq-1})-(\ref{eq-3}), in view of GSD theory, in only one direction in the $(r,t)$-plane may be written as:

\begin{equation}\label{eq-15}
\frac{\partial p}{\partial t}+(u\pm a)\frac{\partial p}{\partial r}\pm a\rho\frac{\partial u}{\partial t}\pm a\rho(u\pm a)\frac{\partial u}{\partial r}+jK_{s}\frac{u}{r}=0,
\end{equation}
It shows the fact that the positive $C_+$ and the negative $C_-$ characteristic curves in the $(r,t)$-plane represent the motion of  diverging and converging shock waves, respectively. 
Now considering the flow properties behind the shock front, the equivalent form of Eq. (\ref{eq-15}) for $C_-$ may be written as: 

\begin{equation}\label{eq-16}
dp-a\rho du+jK_{s}\frac{u}{u-a}\frac{dr}{r}=0, 
\end{equation}
Whitham's characteristic equation (\ref{eq-16}) along with Eqs. (\ref{eq-10})-(\ref{eq-14}) predicts the regime behind the strong converging shock waves in the compressed solid materials. By substituting Eqs. (\ref{eq-10})-(\ref{eq-14}) into Eq.(\ref{eq-16}), we get a first-order ordinary differential equation in $U$ for the converging shock wave as:

\begin{equation}\label{eq-17}
\frac{dU^2}{U^2}=2jN\frac{dr}{r},
\end{equation}
where $N=1+N_1+N_2$,
\[N_1=\frac{3(G_2^2+G^2)}{5(G_m^2+G^2+(3\Gamma_o-2)(G_m^2+1)G)}\left(2+\sqrt{\frac{6(\Gamma_o+1)(G_m^2+G^2)}{2(G_m^2+G^2)+(3\Gamma_o-2)(G_m^2+1)G}}\right), \]
\[N_2=\sqrt{\frac{2\left[2(G_m^2+G^2)+(3\Gamma_o-2)(G_m^2+1)G)\right]}{3(\Gamma_o+1)(G_m^2+G^2)}}.\]
Eq. (\ref{eq-17}) defines the geometrical structure of the shock-front, and on integrating Eq. (\ref{eq-17}), we get the square of shock velocity as: $U^2=K'r^{-2jN}$, where $K'$ is the constant of integration. Thus, the non-dimensional expression for the shock velocity may be written as: 

\begin{equation}\label{eq-18}
\frac{U}{a_o}=Kr^{-jN},
\end{equation}
where $K=\sqrt{K'}/a_o$ is a constant. Eq. (\ref{eq-18}) is the GSD motion rule and is valid only for the strong converging shock waves. It is important to mention that the effects of re-reflected waves or overtaking disturbances are insignificant in the case of strong converging shock waves \cite{Whitham2011}. 
\begin{table}
\caption{Computed values of exponent $N$}
\label{tab:2}
\begin{tabular}
{p{1.0 cm} p{1.5 cm} p{1.5 cm} p{1.5 cm} p{1.5 cm} p{1.5 cm} p{1.5 cm} p{1.5 cm} p{1.5 cm}}
\hline\noalign{\smallskip}
$G_{m}$ & Ti & Brass & Ta  &  Fe & Steel & Al  & Cu \\
 \noalign{\smallskip}\hline\noalign{\smallskip}
0.51 & 4.15229 & 4.04142 & 3.90788 & 3.84774 & 3.76778 & 3.74696 & 3.72127 \\
0.53 & 4.14722 & 4.03481 & 3.89986 & 3.83925 & 3.75886 & 3.73796 & 3.71219 \\
0.55 & 4.14200 & 4.02803 & 3.89165 & 3.83059 & 3.74979 & 3.72882 & 3.70298 \\
0.57 & 4.13663 & 4.02107 & 3.88329 & 3.82179 & 3.74061 & 3.71957 & 3.69367 \\
0.59 & 4.13112 & 4.01396 & 3.87479 & 3.81287 & 3.73132 & 3.71023 & 3.68429 \\
0.61 & 4.12548 & 4.00671 & 3.86615 & 3.80382 & 3.72195 & 3.70081 & 3.67484 \\
0.63 & 4.11970 & 3.99931 & 3.85740 & 3.79469 & 3.71252 & 3.69135 & 3.66535 \\
0.65 & 4.11380 & 3.99178 & 3.84855 & 3.78548 & 3.70305 & 3.68186 & 3.65585 \\
0.67 & 4.10778 & 3.98414 & 3.83962 & 3.77620 & 3.69356 & 3.67235 & 3.64635 \\
0.69 & 4.10164 & 3.97639 & 3.83061 & 3.76689 & 3.68406 & 3.66285 & 3.63686 \\
0.71 & 4.09540 & 3.96853 & 3.82155 & 3.75754 & 3.67458 & 3.65337 & 3.62742 \\
0.73 & 4.08905 & 3.96059 & 3.81245 & 3.74819 & 3.66513 & 3.64394 & 3.61803 \\
0.75 & 4.08261 & 3.95258 & 3.80333 & 3.73884 & 3.65573 & 3.63456 & 3.60871 \\
0.77 & 4.07609 & 3.94449 & 3.79419 & 3.72952 & 3.64639 & 3.62527 & 3.59949 \\
0.79 & 4.06948 & 3.93635 & 3.78507 & 3.72023 & 3.63713 & 3.61606 & 3.59036 \\
 \noalign{\smallskip}\hline
\end{tabular}
\end{table}
Now, the analytical expressions for the flow quantities can be easily written using the expression of shock velocity (\ref{eq-18}) into the Rankine-Hugoniot jump conditions (\ref{eq-10})-(\ref{eq-14}). Thus, the non-dimensional  expressions for the pressure, the mass density, the particle velocity, the temperature, the speed of sound, the adiabatic bulk modulus, and the change-in-entropy behind the strong converging shock front are, respectively: 

\begin{equation}\label{eq-19}
\frac{p}{p_o}=\frac{2K^2N_4}{N_5}r^{-2jN},
\end{equation}

\begin{equation}\label{eq-20}
\frac{\rho}{\rho_o}=\frac{N_5}{N_3},
\end{equation}

\begin{equation}\label{eq-21}
\frac{u}{a_o}=\frac{6K(G_m^2+G^2)}{N_5}r^{-jN},
\end{equation}

\begin{equation}\label{eq-22}
\frac{T}{T_o}=\frac{6K^2\Gamma_o(G_o^2+G^2)N_4}{N_5^2}r^{-2jN},
\end{equation}

\begin{equation}\label{eq-23}
\frac{a}{a_o}=\frac{KN_4}{N_5}\left(\frac{2N_3}{3(\Gamma_o+1)(G_m^2+G^2)}\right)^{1/2} r^{-jN},
\end{equation}

\begin{equation}\label{eq-24}
\frac{K_s}{p_o}=\frac{2K^2N_4}{3(G_m^2+G^2)N_5}r^{-2jN},
\end{equation}

\begin{equation}\label{eq-25}
\frac{\Delta s}{c_v}=ln\left(\frac{2K^2N_4}{N_5}r^{-2jN}\right)-\frac{N_4}{3(G_m^2+G^2)}ln\left(\frac{N_5}{N_3}\right),
\end{equation}
where $N_3=2G_m^2+(3\Gamma_o-2)(G_m^2+1)G+2G^2$, 
\[N_4=5G_m^2+(3\Gamma_o-2)(G_m^2+1)G+5G^2, \] 
\[N_5=8G_m^2+(3\Gamma_o-2)(G_m^2+1)G+8G^2. \]
Thus, the propagation velocity of strong converging cylindrical ($j=1$) or spherical ($j=2$) shock wave, and the distribution of flow quantities in the shock-compressed solid materials can be described using the above expressions (\ref{eq-18})-(\ref{eq-25}).

\section{Results and conclusions}
\label{res}
In the present paper, a one-dimensional analytical solution for strong converging shock waves propagating in a compressed solid material has been found in view of GSD theory. The shock velocity and the properties of shock-compressed materials such as titanium Ti6Al4V, OFHC copper, brass (66\% copper and 34\% zinc), stainless steel 304, tantalum, iron, and aluminum 6061-T6 have been investigated considering both the cases of cylindrical and spherical shock waves. Using Eqs. (\ref{eq-4}) and (\ref{eq-11}), we obtain an expression:

\begin{equation}\label{eq-26}
\left[\frac{2}{3}+\left(\Gamma_o-\frac{2}{3}\right)\frac{G_m^2+1}{G_m^2+G}G\right]\left(G-1\right)=2
\end{equation}
Obviously, the value of $G$ depends on the Gr\"uneisen parameter $\Gamma_o$ and the material parameter $G_m$. The value of Gr\"uneisen parameter $\Gamma_o$ \cite{Bushman1983,Anisimov1985,Steinberg1996} for titanium Ti6Al4V, brass (66\% copper and 34\% zinc), tantalum or tungsten, iron, stainless steel 304, aluminum 6061-T6, and OFHC copper is 1.23, 1.43, 1.67, 1.78, 1.93, 1.97, and 2.02, respectively. It is important to mention that the range of parameter $G_m$ is $0.5<G_m<0.8$. The value of parameter $G$ is computed from Eq. (\ref{eq-26}) and is given in Table \ref{tab:1} for respective materials.

The expressions for the shock velocity (\ref{eq-18}) and the flow quantities (\ref{eq-19})-(\ref{eq-25}) are  primarily the same for cylindrical and spherical symmetries of the converging shocks, except for a geometrical factor $j$ that differs between the two cases. The exponent $N$ depends on $\Gamma_o$ and $G_m$. Table \ref{tab:2} shows the calculated values of the exponent $N$ for the respective solid materials. The exponent $N$ decreases linearly with an increase in the value of the Gr\"uneisen parameter $\Gamma_o$ and material parameter $G_m$.

\begin{figure}   
   \begin{minipage}{0.4\textwidth}
     \centering
     \includegraphics[width=.8\linewidth, trim=0 .1cm .1cm 0, clip]{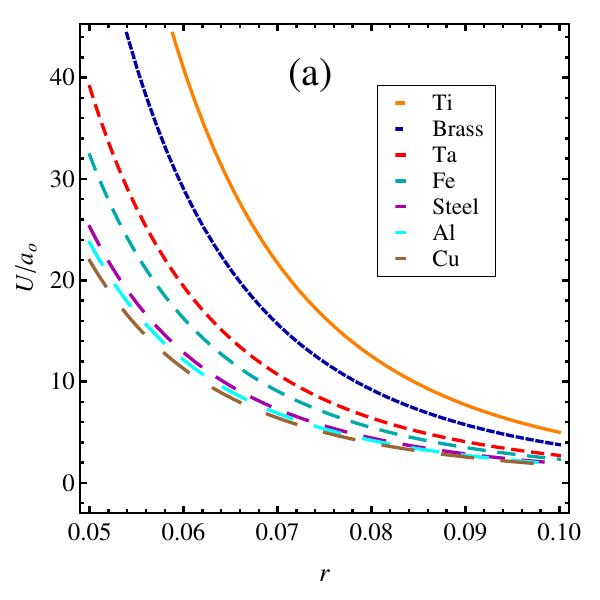}
       \end{minipage}\hfill
   \begin{minipage}{0.4\textwidth}
     \centering
     \includegraphics[width=.8\linewidth, trim=0 .1cm .1cm 0, clip]{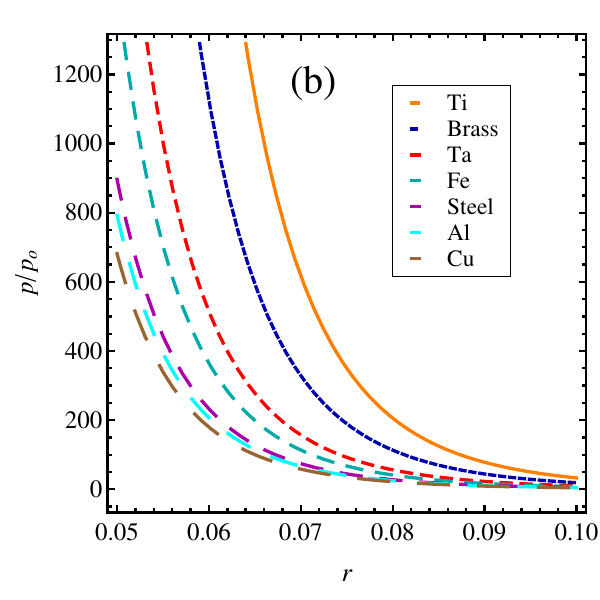}
     \end{minipage}
     \begin{minipage}{0.4\textwidth}
     \centering
     \includegraphics[width=.8\linewidth, trim=0 .1cm .1cm 0, clip]{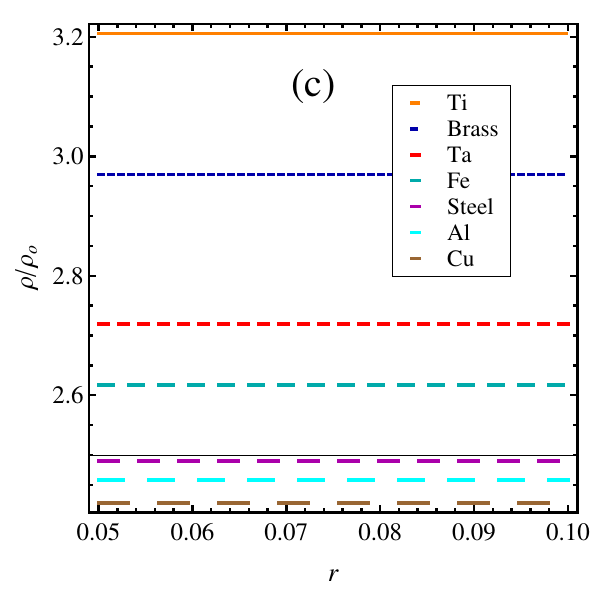}
       \end{minipage}\hfill
   \begin{minipage}{0.4\textwidth}
     \centering
     \includegraphics[width=.8\linewidth, trim=0 .1cm .1cm 0, clip]{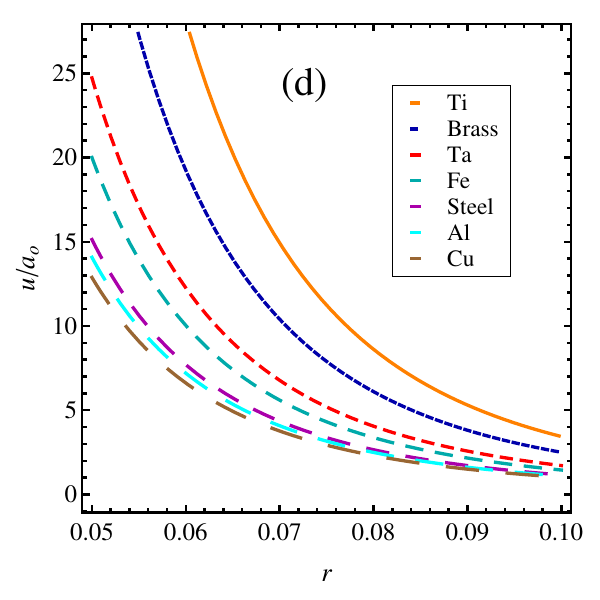}
     \end{minipage}
     \begin{minipage}{0.4\textwidth}
     \centering
     \includegraphics[width=.8\linewidth, trim=0 .1cm .1cm 0, clip]{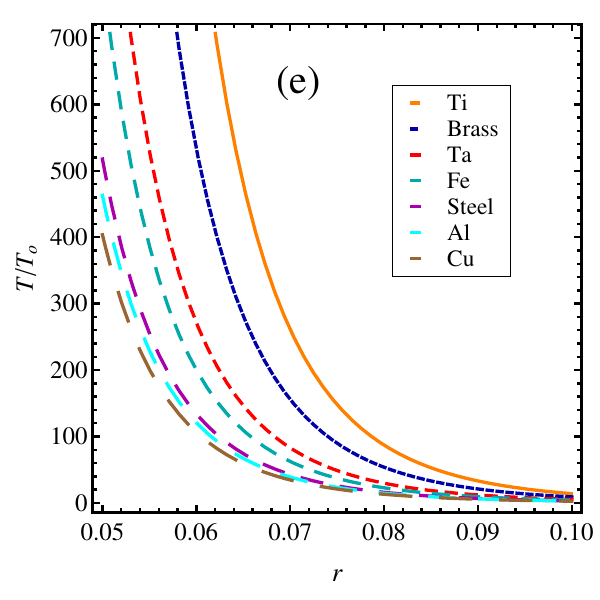}
       \end{minipage}\hfill
   \begin{minipage}{0.4\textwidth}
     \centering
     \includegraphics[width=.8\linewidth, trim=0 .1cm .1cm 0, clip]{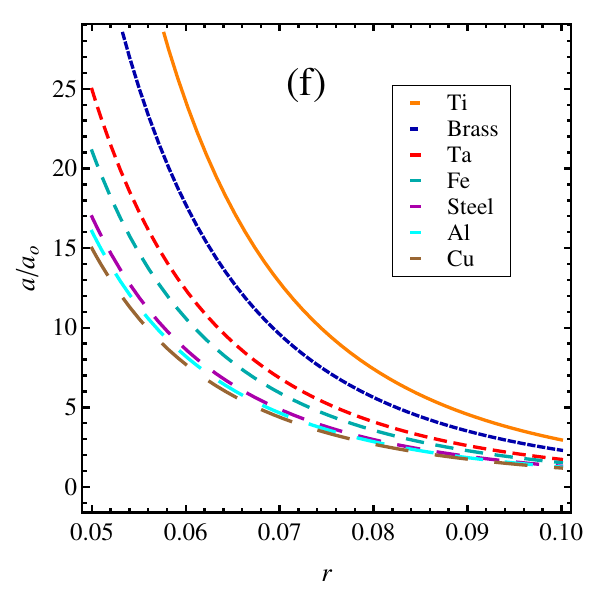}
     \end{minipage}
     \begin{minipage}{0.4\textwidth}
     \centering
     \includegraphics[width=.8\linewidth, trim=0 .1cm .05cm 0, clip]{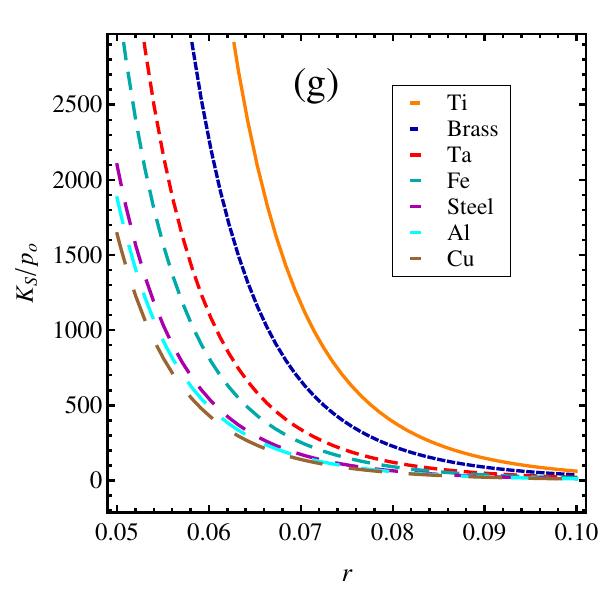}
       \end{minipage}\hfill
   \begin{minipage}{0.4\textwidth}
     \centering
     \includegraphics[width=.8\linewidth, trim=0 .05cm .05cm 0, clip]{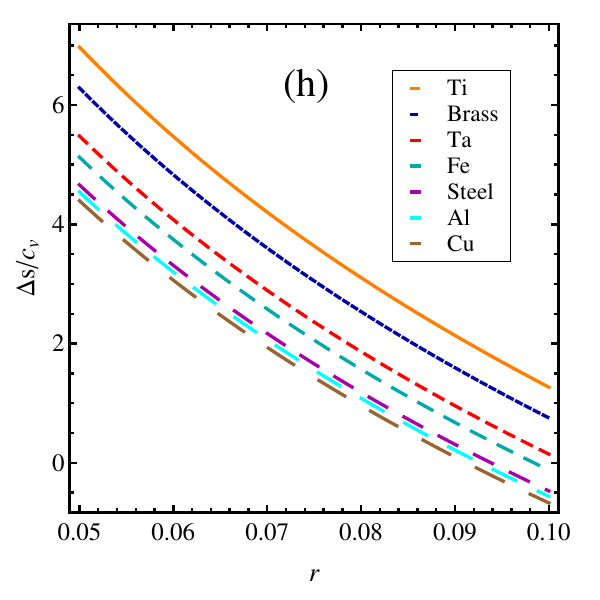}
     \end{minipage}
\caption{The variations of non-dimensional (a) shock velocity $U/a_{o}$, (b) pressure $p/p_{o}$, (c) mass density $\rho/\rho_{o}$, (d) particle velocity $u/a_{o}$, (e) temperature $T/T_{o}$, (f) speed of sound $a/a_{o}$, (g) adiabatic bulk modulus $K_s/p_o$, and (h) change-in-entropy $\Delta s/c_{v}$ with distance $r$ for the case of a strong converging cylindrical shock wave}\label{figure1a_1h}
\end{figure}

\begin{figure}  
   \begin{minipage}{0.4\textwidth}
     \centering
     \includegraphics[width=.8\linewidth, trim=0 .1cm .1cm 0, clip]{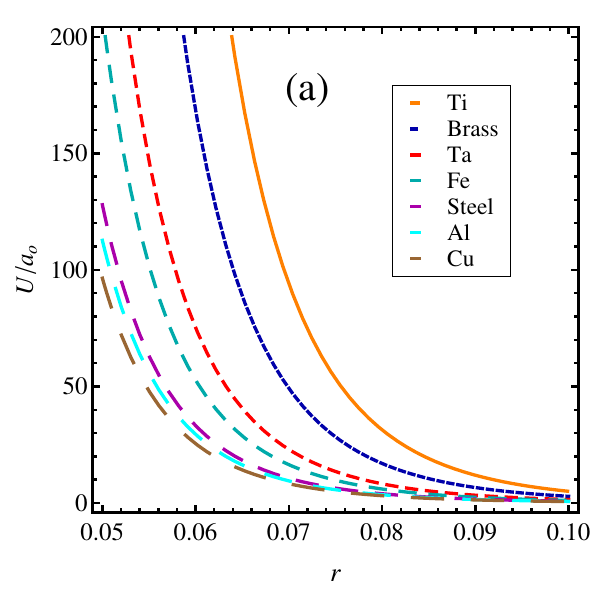}
       \end{minipage}\hfill
   \begin{minipage}{0.4\textwidth}
     \centering
     \includegraphics[width=.8\linewidth, trim=0 .1cm .1cm 0, clip]{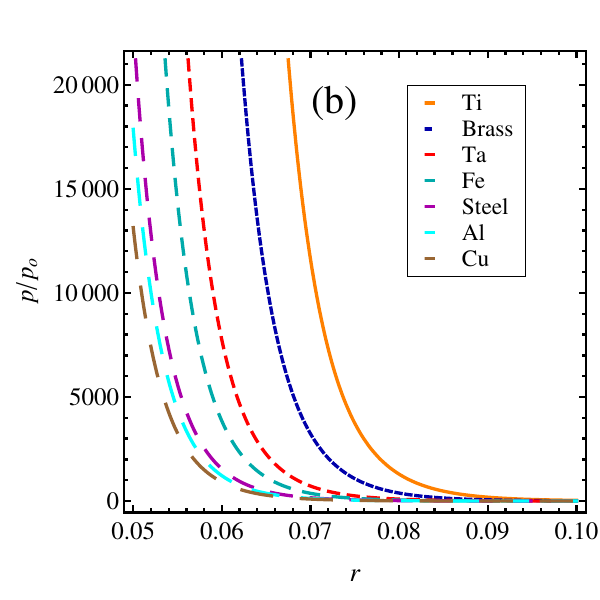}
     \end{minipage}
     \begin{minipage}{0.4\textwidth}
     \centering
     \includegraphics[width=.8\linewidth, trim=0 .1cm .1cm 0, clip]{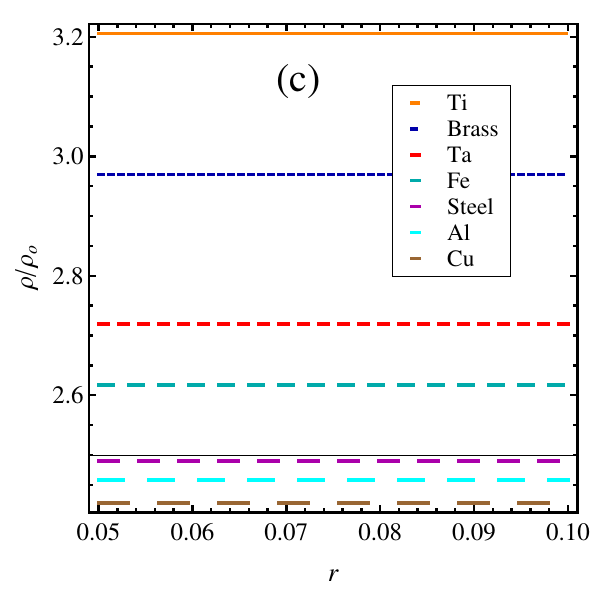}
       \end{minipage}\hfill
   \begin{minipage}{0.4\textwidth}
     \centering
     \includegraphics[width=.8\linewidth, trim=0 .1cm .1cm 0, clip]{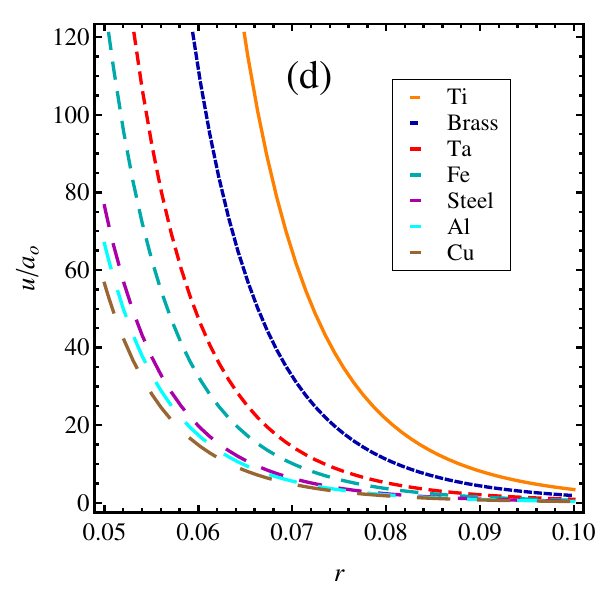}
     \end{minipage}
     \begin{minipage}{0.4\textwidth}
     \centering
     \includegraphics[width=.8\linewidth, trim=0 .1cm .1cm 0, clip]{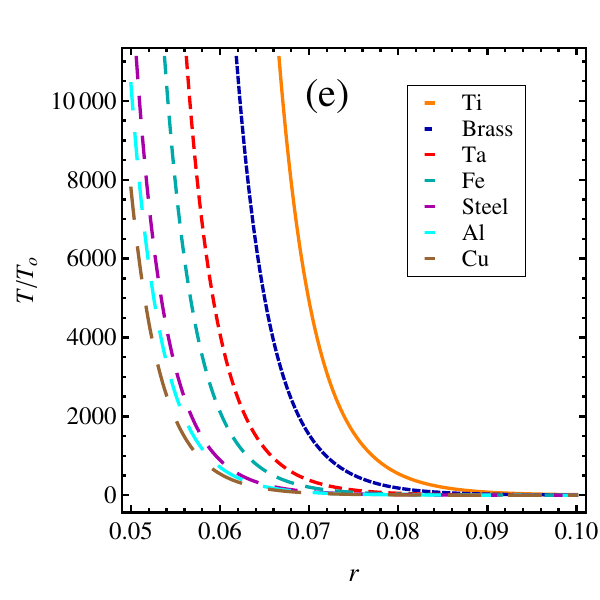}
       \end{minipage}\hfill
   \begin{minipage}{0.4\textwidth}
     \centering
     \includegraphics[width=.8\linewidth, trim=0 .1cm .1cm 0, clip]{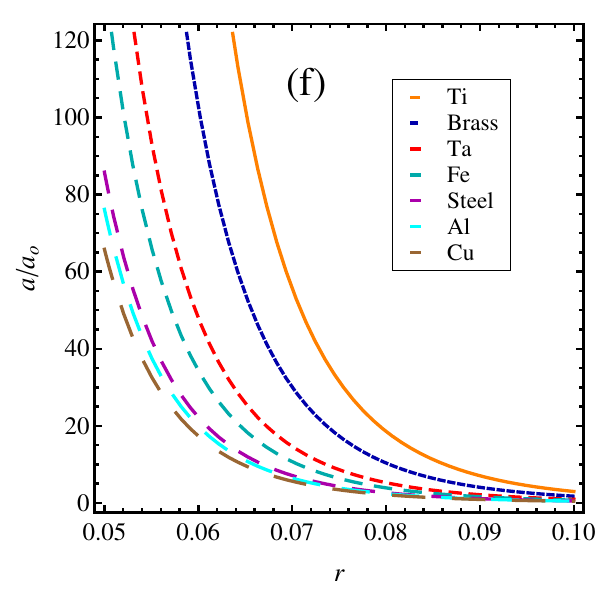}
     \end{minipage}
     \begin{minipage}{0.4\textwidth}
     \centering
     \includegraphics[width=.8\linewidth, trim=0 .1cm .05cm 0, clip]{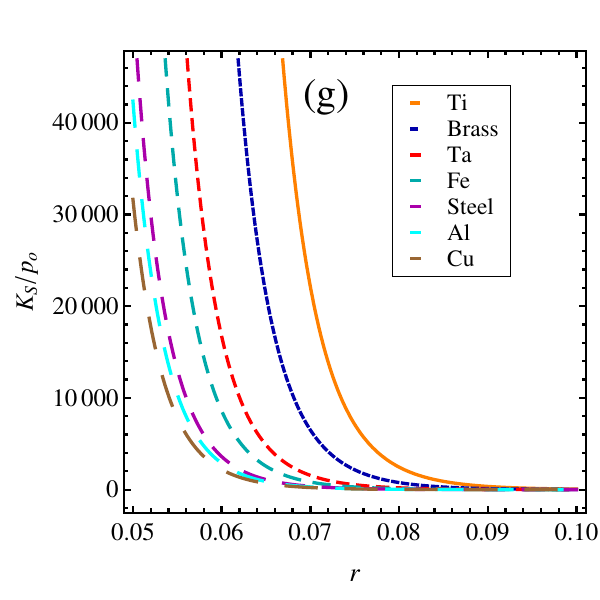}
       \end{minipage}\hfill
   \begin{minipage}{0.4\textwidth}
     \centering
     \includegraphics[width=.8\linewidth, trim=0 .05cm .05cm 0, clip]{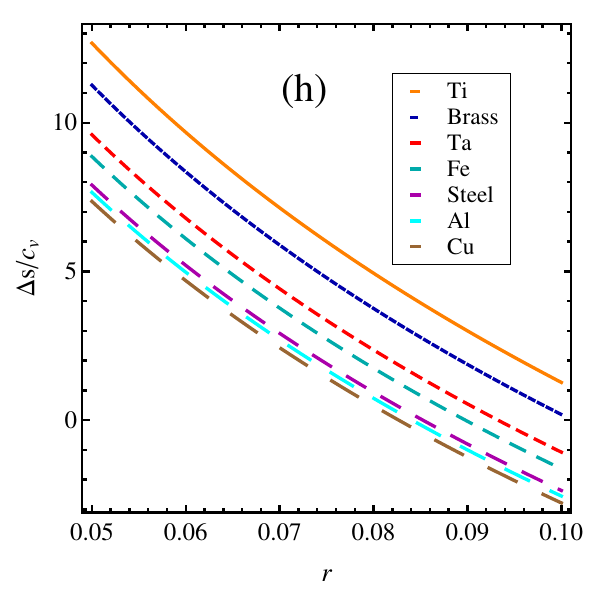}
     \end{minipage}
\caption{The variations of non-dimensional (a) shock velocity $U/a_{o}$, (b) pressure $p/p_{o}$, (c) mass density $\rho/\rho_{o}$, (d) particle velocity $u/a_{o}$, (e) temperature $T/T_{o}$, (f) speed of sound $a/a_{o}$, (g) adiabatic bulk modulus $K_s/p_o$, and (h) change-in-entropy $\Delta s/c_{v}$ with distance $r$ for the case of a strong converging spherical shock wave}\label{figure2a_2h}
\end{figure}

The analytical expressions (\ref{eq-18})-(\ref{eq-25}) have been computed by taking  $K=0.000 384 743$ and $K=2.96055\times10^{-8}$ for cylindrical and spherical shock, respectively. Here, taking $U=5a_o$  at $r=0.1$ for $j=$ 1 or 2, $\Gamma_o=1.23$ and $G_m=0.65$, the value of constant $K$ is determined using Eq. (\ref{eq-18}). The variations of (a) shock velocity, (b) pressure, (c) mass density, (d) particle velocity, (e) temperature, (f) speed of sound, (g) adiabatic bulk modulus, and (h) change-in-entropy with propagation distance $r$ taking $G_m=0.65$ are shown in Figs. \ref{figure1a_1h} and \ref{figure2a_2h} for respective shock-compressed materials. As the shock wave moves towards the centre of convergence, the shock velocity increases, and an increase is also observed in the pressure, the particle velocity, the temperature, the speed of sound, the bulk modulus, and the change-in-entropy. It is seen in Figs. \ref{figure1a_1h}(c)  and \ref{figure2a_2h}(c) that  the mass density remains unchanged with the propagation distance $r$ for both the symmetries of cylindrical and spherical shocks. A decrease is observed in the shock velocity with the Gr\"uneisen parameter $\Gamma_o$, and it is maximum in the compressed titanium Ti6Al4V but minimum in the compressed OFHC copper; see Tables \ref{tab:3} and \ref{tab:4}. There is a decrease in the pressure, the mass density, the particle velocity, the temperature, and the speed of sound with the Gr\"uneisen parameter. The numerical values of these quantities are higher in the compressed titanium Ti6Al4V than in the compressed OFHC copper. The adiabatic bulk modulus of materials after the passage of shock comes down with an increasing value of Gr\"uneisen parameter. Similar behaviour is observed in the case of change-in-entropy in shock-compressed solids. 
\begin{table}
\caption{The shock velocity and flow quantities for the cylindrical symmetry of the shock wave}
\label{tab:3}
\begin{tabular}
{p{1.0 cm} p{0.8 cm} p{1.5 cm} p{1.5 cm} p{1.5 cm} p{1.5 cm} p{1.5 cm} p{1.5 cm} p{1.5 cm}}
\hline\noalign{\smallskip}
 & $G_m$ & $ Ti$ & $Brass$ & $Ta$ & $Fe$ & $Steel$ & $Al$ & $Cu$\\
 \noalign{\smallskip}\hline\noalign{\smallskip}
 $U/a_o$ & 0.51 & 97.1454 & 69.6915 & 46.7136 & 39.0116 & 30.7023 & 28.8456 & 26.7086 \\
 & 0.58 & 91.9376 & 64.8798 & 42.8491 & 35.6165 & 27.9115 & 26.2046 & 24.2469 \\
 & 0.65 & 86.5659 & 60.0623 & 39.1067 & 32.3734 & 25.2903 & 23.7342 & 21.9555 \\
 & 0.72 & 81.1514 & 55.3606 & 35.5803 & 29.3605 & 22.8960 & 21.4868 & 19.8809 \\
 & 0.79 & 75.8024 & 50.8725 & 32.3336 & 26.6254 & 20.7582 & 19.4881 & 18.0442 \\
&\\
 $p/p_o$ & 0.51 & 12316.5 & 6446.39 & 2957.10 & 2082.29 & 1306.71 & 1157.45 & 996.599 \\
 & 0.58 & 11058.8 & 5605.99 & 2499.11 & 1744.04 & 1085.71 & 960.419 & 825.946 \\
 & 0.65 & 9831.32 & 4822.44 & 2091.69 & 1448.46 & 896.475 & 792.475 & 681.254 \\
 & 0.72 & 8665.99 & 4113.70 & 1740.41 & 1198.03 & 739.192 & 653.478 & 562.074 \\
 & 0.79 & 7585.80 & 3488.85 & 1445.07 & 990.945 & 611.368 & 540.932 & 465.966 \\
&\\
 $\rho/\rho_o$ & 0.51 & 3.27758 & 3.05567 & 2.81588 & 2.71579 & 2.58913 & 2.55723 & 2.51846 \\
 & 0.58 & 3.24315 & 3.01402 & 2.76906 & 2.66776 & 2.54046 & 2.50856 & 2.46988 \\
 & 0.65 & 3.20562 & 2.96919 & 2.71951 & 2.61733 & 2.48990 & 2.45815 & 2.41974 \\
 & 0.72 & 3.16546 & 2.92191 & 2.66823 & 2.56560 & 2.43864 & 2.40719 & 2.36924 \\
 & 0.79 & 3.12316 & 2.87292 & 2.61619 & 2.51360 & 2.38775 & 2.35676 & 2.31945 \\
&\\
 $u/a_o$ & 0.51 & 67.5060 & 46.8842 & 30.1242 & 24.6468 & 18.8442 & 17.5656 & 16.1035 \\
 & 0.58 & 63.5893 & 43.3538 & 27.3749 & 22.2658 & 16.9247 & 15.7586 & 14.4299 \\
 & 0.65 & 59.5615 & 39.8338 & 24.7266 & 20.0046 & 15.1331 & 14.0789 & 12.882 \\
 & 0.72 & 55.5148 & 36.4139 & 22.2455 & 17.9166 & 13.5072 & 12.5607 & 11.4896 \\
 & 0.79 & 51.5313 & 33.1649 & 19.9746 & 16.0329 & 12.0646 & 11.2191 & 10.2647 \\
&\\
 $T/T_o$ & 0.51 & 5263.62 & 3100.77 & 1592.30 & 1170.84 & 773.946 & 694.260 & 606.889 \\
 & 0.58 & 4704.07 & 2678.40 & 1333.16 & 970.359 & 635.302 & 568.899 & 496.454 \\
 & 0.65 & 4160.12 & 2286.77 & 1104.33 & 796.594 & 517.656 & 463.037 & 403.711 \\
 & 0.72 & 3645.92 & 1934.66 & 908.596 & 650.655 & 420.813 & 376.279 & 328.084 \\
 & 0.79 & 3171.50 & 1626.23 & 745.419 & 531.074 & 342.888 & 306.737 & 267.722 \\
&\\
 $a/a_o$ & 0.51 & 56.2572 & 41.3863 & 28.7491 & 24.4396 & 19.7239 & 18.6581 & 17.4247 \\
 & 0.58 & 53.7815 & 39.0578 & 26.8359 & 22.7413 & 18.3093 & 17.3150 & 16.1678 \\
 & 0.65 & 51.2091 & 36.7044 & 24.9625 & 21.1006 & 16.9657 & 16.0447 & 14.9851 \\
 & 0.72 & 48.5954 & 34.3845 & 23.1771 & 19.5589 & 15.7249 & 14.8766 & 13.9032 \\
 & 0.79 & 45.9913 & 32.1467 & 21.5146 & 18.1436 & 14.6057 & 13.8275 & 12.9365 \\
&\\
 $K_s/p_o$ & 0.51 & 23132.0 & 12718.2 & 6214.05 & 4509.51 & 2951.26 & 2644.00 & 2309.25 \\
 & 0.58 & 20918.9 & 11172.9 & 5324.45 & 3835.51 & 2495.31 & 2233.71 & 1949.78 \\
 & 0.65 & 18746.1 & 9720.33 & 4524.58 & 3239.63 & 2099.88 & 1879.43 & 1640.94 \\
 & 0.72 & 16669.8 & 8394.56 & 3826.95 & 2728.48 & 1766.82 & 1582.25 & 1383.08 \\
 & 0.79 & 14731.6 & 7214.42 & 3233.32 & 2300.33 & 1492.46 & 1338.32 & 1172.27 \\
&\\
 $\Delta s/c_v$ & 0.51 & 7.18917 & 6.56753 & 5.81645 & 5.47757 & 5.02666 & 4.90916 & 4.76414 \\
 & 0.58 & 7.08542 & 6.43272 & 5.65371 & 5.30600 & 4.84717 & 4.72834 & 4.58210 \\
 & 0.65 & 6.97212 & 6.28743 & 5.48163 & 5.12629 & 4.66166 & 4.54212 & 4.39546 \\
 & 0.72 & 6.85061 & 6.13404 & 5.30387 & 4.94263 & 4.47483 & 4.35532 & 4.20914 \\
 & 0.79 & 6.72241 & 5.97506 & 5.12409 & 4.75903 & 4.29101 & 4.17228 & 4.02751 \\
\noalign{\smallskip}\hline
\end{tabular}
\end{table}

\begin{table}
\caption{The shock velocity and flow quantities for the spherical symmetry of the shock wave}
\label{tab:4}
\begin{tabular}
{p{1.0 cm} p{0.8 cm} p{1.5 cm} p{1.5 cm} p{1.5 cm} p{1.5 cm} p{1.5 cm} p{1.5 cm} p{1.5 cm}}
\hline\noalign{\smallskip}
 & $G_m$ & $ Ti$ & $Brass$ & $Ta$ & $Fe$ & $Steel$ & $Al$ & $Cu$\\
 \noalign{\smallskip}\hline\noalign{\smallskip}
 $U/a_o$ & 0.51 & 1887.44 & 971.382 & 436.431 & 304.380 & 188.527 & 166.414 & 142.670 \\
 & 0.58 & 1690.50 & 841.877 & 367.210 & 253.706 & 155.811 & 137.337 & 117.583 \\
 & 0.65 & 1498.73 & 721.495 & 305.867 & 209.608 & 127.920 & 112.663 & 96.4084 \\
 & 0.72 & 1317.11 & 612.960 & 253.191 & 172.407 & 104.845 & 92.3369 & 79.0498 \\
 & 0.79 & 1149.20 & 517.603 & 209.093 & 141.783 & 86.1808 & 75.9569 & 65.1183 \\
&\\
 $p/p_o$ & 0.51 & $4.6\times 10^6$ & $1.3\times 10^6$ & 258115. & 126762. & 49269.8 & 38523.3 & 28436.9 \\
 & 0.58 & $3.7\times 10^6$ & 943910. & 183539. & 88494.7 & 33833.1 & 26380.1 & 19423.4 \\
 & 0.65 & $2.9\times 10^6$ & 695874. & 127956. & 60721.6 & 22935.4 & 17856.5 & 13135.7 \\
 & 0.72 & $2.3\times 10^6$ & 504307. & 88131.4 & 41310.0 & 15500.1 & 12068.0 & 8886.36 \\
 & 0.79 & $1.7\times 10^6$ & 361167. & 60430.9 & 28099.8 & 10537.6 & 8217.50 & 6068.58 \\
&\\
 $\rho/\rho_o$ & 0.51 & 3.27758 & 3.05567 & 2.81588 & 2.71579 & 2.58913 & 2.55723 & 2.51846 \\
 & 0.58 & 3.24315 & 3.01402 & 2.76906 & 2.66776 & 2.54046 & 2.50856 & 2.46988 \\
 & 0.65 & 3.20562 & 2.96919 & 2.71951 & 2.61733 & 2.48990 & 2.45815 & 2.41974 \\
 & 0.72 & 3.16546 & 2.92191 & 2.66823 & 2.56560 & 2.43864 & 2.40719 & 2.36924 \\
 & 0.79 & 3.12316 & 2.87292 & 2.61619 & 2.51360 & 2.38775 & 2.35676 & 2.31945 \\
&\\
 $u/a_o$ & 0.51 & 1311.58 & 653.487 & 281.442 & 192.302 & 115.712 & 101.338 & 86.0202 \\
 & 0.58 & 1169.25 & 562.557 & 234.598 & 158.606 & 94.4790 & 82.5895 & 69.9759 \\
 & 0.65 & 1031.20 & 478.501 & 193.395 & 129.523 & 76.5443 & 66.8304 & 56.5659 \\
 & 0.72 & 901.022 & 403.179 & 158.300 & 105.208 & 61.8519 & 53.9782 & 45.6847 \\
 & 0.79 & 781.239 & 337.436 & 129.170 & 85.3766 & 50.0880 & 43.7276 & 37.0435 \\
&\\
 $T/T_o$ & 0.51 & $2.0\times 10^6$ & 602406. & 138986. & 71276.4 & 29181.9 & 23106.9 & 17317.0 \\
 & 0.58 & $1.6\times 10^6$ & 450977. & 97909.7 & 49237.3 & 19797.4 & 15626.1 & 11674.9 \\
 & 0.65 & $1.2\times 10^6$ & 329979. & 67555.4 & 33394.4 & 13243.7 & 10433.4 & 7784.22 \\
 & 0.72 & 960414. & 237174. & 46009.7 & 22435.5 & 8824.04 & 6948.89 & 5187.00 \\
 & 0.79 & 728938. & 168349. & 31172.3 & 15059.4 & 5910.08 & 4659.76 & 3486.72 \\
&\\
 $a/a_o$ & 0.51 & 1093.03 & 576.855 & 268.595 & 190.686 & 121.114 & 107.641 & 93.0777 \\
 & 0.58 & 988.909 & 506.812 & 229.979 & 161.993 & 102.208 & 90.7467 & 78.4039 \\
 & 0.65 & 886.593 & 440.910 & 195.240 & 136.620 & 85.8135 & 76.1615 & 65.8009 \\
 & 0.72 & 788.717 & 380.709 & 164.930 & 114.852 & 72.0074 & 63.9302 & 55.2816 \\
 & 0.79 & 697.250 & 327.077 & 139.129 & 96.6164 & 60.6377 & 53.8943 & 46.6857 \\
&\\
 $K_s/p_o$ & 0.51 & $8.7\times 10^6$ & $2.5\times 10^6$ & 542401. & 274521. & 111278. & 87999.8 & 65892.0 \\
 & 0.58 & $7.1\times 10^6$ & $1.9\times 10^6$ & 391038. & 194619. & 77759.3 & 61354.0 & 45851.9 \\
 & 0.65 & $5.6\times 10^6$ & $1.4\times 10^6$ & 276783. & 135810. & 53723.1 & 42348.4 & 31640.2 \\
 & 0.72 & $4.4\times 10^6$ & $1.0\times 10^6$ & 193790. & 94082.1 & 37048.4 & 29220.0 & 21866.4 \\
 & 0.79 & $3.4\times 10^6$ & 746841. & 135213. & 65229.4 & 25724.2 & 20330.9 & 15267.2 \\
&\\
 $\Delta s/c_v$ & 0.51 & 13.1227 & 11.8368 & 10.2856 & 9.58641 & 8.65646 & 8.41420 & 8.11524 \\
 & 0.58 & 12.9088 & 11.5589 & 9.95021 & 9.23274 & 8.28638 & 8.04134 & 7.73980 \\
 & 0.65 & 12.6751 & 11.2593 & 9.59534 & 8.86209 & 7.90362 & 7.65708 & 7.35461 \\
 & 0.72 & 12.4244 & 10.9429 & 9.22857 & 8.48305 & 7.51788 & 7.27132 & 6.96978 \\
 & 0.79 & 12.1598 & 10.6148 & 8.85743 & 8.10389 & 7.13802 & 6.89301 & 6.59428 \\
\noalign{\smallskip}\hline
\end{tabular}
\end{table}

It is obvious from Tables \ref{tab:3} and \ref{tab:4} that the shock velocity and the flow quantities behind the strong converging shock front decrease with an increase in the value of the material parameter $G_m$. Thus, in general, the numerical values of the shock velocity and other quantities are greater in the compressed titanium Ti6Al4V than in the compressed OFHC copper. The trends of fluctuations in the shock velocity and the material properties are similar in the shock-compressed aluminum 6061-T6, titanium Ti6Al4V, iron, OFHC copper, brass (66\% copper and 34\% zinc), stainless steel 304, and tantalum.

The distribution of the mass density behind the converging cylindrical or spherical shock wave is independent of the propagation distance $r$; see Eq.(\ref{eq-21}). Therefore, the mass density behind the shock remains unchanged with the propagation distance, but it is notable that its numerical value with $G_m=0.65$ is 3.20562, 2.96919, 2.71951, 2.61733, 2.48990, 2.45815, and 2.41974, respectively, for titanium Ti6Al4V, brass (66\% copper and 34\% zinc), tantalum, iron, stainless steel 304, aluminum 6061-T6, and OFHC copper. The mass density behind the cylindrical or spherical shock front decreases with an increase in the value of the Gr\"uneisen parameter $\Gamma_o$ and material parameter $G_m$. 

Obviously, the trends of fluctuations in the shock velocity and the flow quantities are similar for cylindrical and spherical symmetries of strong converging shock waves; however, the rates of increase or decrease, i.e., the numerical values of these quantities, are larger for spherical shocks. At the focal regions, temperatures in the range of 13000-34000 K have been observed in the case of cylindrical or spherical converging shock waves using spectroscopic techniques; thus, our results are in good agreement with the experimental observations \cite{Matsuo1985,Liverts2016}. Moreover, the findings of the present study might also prove to be a valuable reference for material scientists and engineers. The methodology and analysis presented in this paper may be used to investigate the propagation of shock waves in nanofluids.


\end{document}